\documentstyle[12pt]{article}
\newcommand{\eeqn}[1]{\label{#1}\end{equation}}
\newcommand{\eean}[1]{\label{#1}\end{eqnarray}}
\def\npb#1#2#3{    {\it Nucl. Phys. }{\bf B #1} (19#2) #3}
\def\npbps#1#2#3{    {\it Nucl. Phys. }(Proc. Suppl.){\bf B #1} (19#2) #3}
\def\plb#1#2#3{    {\it Phys. Lett. }{\bf B #1} (19#2) #3}
\def\prd#1#2#3{    {\it Phys. Rev. }{\bf D #1} (19#2) #3}

\def\prl#1#2#3{    {\it Phys. Rev. Lett. }{\bf #1} (19#2) #3}

\def\zpc#1#2#3{    {\it Zeit. f\"ur Physik }{\bf C #1} (19#2) #3}
\def\mpla#1#2#3{   {\it Mod. Phys. Lett. }{\bf A #1} (19#2) #3}

\def\beq{\begin{equation}}
\def\eeq{\end{equation}}
\def\bea{\begin{eqnarray}}
\def\eea{\end{eqnarray}}
\def\ba{\begin{array}}
\def\ea{\end{array}}
\def\ltap{\ \raisebox{-.4ex}{\rlap{$\sim$}} \raisebox{.4ex}{$<$}\ }
\def\gtap{\ \raisebox{-.4ex}{\rlap{$\sim$}} \raisebox{.4ex}{$>$}\ }
\def\eq#1{{eq. (\ref{#1})}}
\def\eqs#1#2{{eqs. (\ref{#1}--\ref{#2})}}

\def\GeV{\mathop{\rm GeV}}
\def\eV{\mathop{\rm eV}}
\def\max{\mathop{\rm max}}

\def\etal{{\it et al.}}
\def\eg{{\it e.g. }}
\def\ie{{\it i.e. }}
\begin{document}

\begin{titlepage}
\vspace*{-1.5cm}
\begin{center}
\hfill SISSA 63/94/EP\\
\hfill IC/94/102
\\[1ex]  \hfill May, 1994

\vspace{4ex}
{\Large \bf Neutrino Masses  and $b-\tau$ unification in the
Supersymmetric Standard Model}

\vspace{3ex} {\bf  Francesco Vissani$^{a,b}$ and Alexei Yu. Smirnov$^{c,d}$
}

{\it
\vspace{1ex} ${}^a$ International School for Advanced Studies, SISSA
%\\[-1ex]
\\[-1ex] Via Beirut 2-4, I-34013 Trieste, Italy
}

{\it
\vspace{1ex}  ${}^b$ Istituto Nazionale di Fisica Nucleare, INFN
\\[-1ex] Sezione di Trieste, c/o SISSA
}

{\it
\vspace{1ex}  ${}^c$ International Center for Theoretical Physics, ICTP
%\\[-1ex]
\\[-1ex] Via Costiera 11, I-34013 Trieste, Italy
}

{\it
\vspace{1ex}  ${}^d$ Institute for Nuclear Research,
\\[-1ex]Russian Academy of Sciences,
\\[-1ex] 117312 Moscow,  Russia
}

\vspace{6ex}
{ ABSTRACT}
\end{center}
\begin{quotation}

There are several indications that the Majorana masses of the
right-handed neutrino components, $M_R,$
are at the intermediate scale: $M_R\sim (10^{10}-10^{12})\ \GeV$ or
even lighter.
The renormalization effects due to large Yukawa
couplings of neutrinos from region of momenta $M_R\ltap
q\ltap M_G$ are studied in the  supersymmetric standard model.
It is shown that neutrino renormalization effect can increase
the $m_b/m_\tau$ ratio up to $(10\div 15)\%.$
This strongly disfavours $m_b-m_\tau$ unification for low values of
$\tan\beta < 10$ especially at  large values of $\alpha_s.$
Lower bounds on $M_R$ and $\tan\beta$ from the $b-\tau$
unification condition were found.
The implications of the results to the see-saw mechanism
of the neutrino mass
generation  are discussed.

\end{quotation}
\end{titlepage}
\vfill\eject

\noindent{\large 1. Introduction.}

The unification of fermion masses, and in particular the masses
of $b-$quark and
$\tau-$lepton \cite{btau}  at the scale $M_G$, where
the gauge coupling unify,
\beq
m_b(M_G)=m_\tau(M_G)
\eeqn{mass unif}
is one of crucial issues  of the  Grand Unification (GU in the
following).
With the standard model particle content
neither $b-\tau$ unification nor  gauge couplings
unification happen to be realized \cite{florida}.
In contrast,
in the minimal supersymmetric extension of the standard model
there is a successful
gauge coupling  unification \cite{gcc}; also the
$b-\tau$ unification is possible, but at the large
Yukawa coupling  of top (and/or of bottom)  quark only, so that
the renormalization effect due to the Yukawa interaction
appreciably suppresses the value of $m_b$ at low scale
\cite{btau susy}.

In previous studies it was suggested
that the right-handed (RH) neutrino components (if they exist)
are at the $M_G$ or higher scale,
so that the neutrinos do not contribute to the
renormalization \cite{decoupled nu tau}.
However, there are several indications that the Majorana
masses of the  RH neutrinos, $M_R$, are much smaller than the GU scale:
$M_R \ll M_{G}$, and consequently the neutrino renormalization
effects may be important. Indeed, studies of the
solar,  atmospheric, as well as
relic (hot dark matter, large scale structure of the universe)
neutrinos
give some hints of the existence of nonzero
neutrino masses \cite{AYS}.
The required values of masses can be naturally generated by
the see-saw mechanism \cite{see-saw} if the
Majorana masses of the RH neutrinos  are at the intermediate
scale: $M_R \sim (10^{10} - 10^{12})\ \GeV$. In particular, for the
tau neutrino to be in the cosmologically interesting domain,
$m_\nu\sim 3-10\ \eV$,  one needs
\beq
M_R\ltap m_t^2/m_\nu \sim 10^{11}~~ {\rm GeV}.
\eeqn{RH bound}

The same scale of masses is required by  mechanism of the  baryon
asymmetry generation based on the decay of the RH neutrinos \cite{Fuk}.
Much lower masses: $M_R < 10^7\ \GeV$, are implied by the
Primordial nucleosynthesis in
the  supersymmetric models with
spontaneous violation of lepton number \cite{Moh}.

The RH neutrinos decouple at $q < M_R$, and if $M_R \ll M_G$, one should
take into account the renormalization effect due to the Yukawa
interactions of neutrinos from the region of momentum
$M_R\ltap q\ltap M_G$. Moreover, in a wide class of the GU theories the
Yukawa couplings of the  neutrinos and  up-quarks are of the same order
(or even equal at the GU scale). Therefore,  at least for the
tau neutrino one can expect large Yukawa coupling, and
consequently, large neutrino renormalization effect.

In this paper we study the renormalization effects of neutrinos on the
ratio $m_b/m_\tau$ as well as  on the
the neutrino masses in the supersymmetric
model.
As the first step the problem is worked out mainly in one loop
approximation and the threshold effects of the new particles at GU scale
as well as at electroweak scales are neglected.

\noindent{\large 2. Renormalization group equations with  RH Neutrinos}

Let us extend  the minimal supersymmetric standard model
by introducing the three right-handed
neutrino superfields $\nu^c_i$ (one for
each family),  and adding to the (matter parity conserving)
superpotential
the terms with neutrino masses and the Yukawa interactions:
\beq
W=-Q Y_u^* U^c H_2+Q Y_d^* D^c H_1+L Y_e^* E^c H_1
- L Y_\nu^* \nu^c H_2
+\mu H_1 H_2
+\frac{1}{2}\nu^c M \nu^c.
\eeqn{superpotential}
Here $Q, L,U^c,D^c,E^c,\nu^c$ are the matter superfields, $H_1$ and
$H_2$ are the  Higgs superfield doublets, and $Y_i$ ($i = u, d, e,$ $\nu$)
are the 3$\times$3 matrices of the  Yukawa couplings.
Taking into account the
the successful unification of gauge couplings in MSSM
we will assume that there is no new particles apart from
$\nu^c$ up to the Grand Unification  scale.
The right-handed neutrino  do not influence the
gauge coupling constants evolution at 1-loop level;
the 2-loop effect is estimated to be small.

We will suggest first that the effect
of flavour mixing is negligible, and there is a hierarchy of the
Yukawa couplings,  so that  the renormalization
effects of particles from
the  third  family is important only. (We will comment
on the effect
of mixing in sect. 3).

It is convenient to write the renormalization
group equation (RGE) for
the couplings:
\beq
\alpha_x={|Y_x|^2 \over 4 \pi}\ \ \ \ \ \ \ \
{\mbox{($x=t,b,\tau,\nu_{\tau}$)}},
\eeqn{scalar alpha}
being the analogous of the gauge coupling constants. In terms of
$\alpha_x$ and the parameter $\tan \beta$, where
$\tan\beta \equiv v_2/v_1,$ and
$v_1$ and $v_2$  are the vacuum expectation values
of the Higgs doublets $H_1,\ H_2,$
the masses of the quarks and  leptons
at the $Z^0$-mass scale, $M_Z$, can be written as:
\beq
\begin{array}{lcl}
m_{t,\nu_{\tau}}(M_Z)& = &\sqrt{4\pi\alpha_{t,\nu_{\tau}}(M_Z)}
\frac{\tan\beta} {\sqrt{1+\tan^2\beta}} v\\
m_{b,\tau}(M_Z)& = &\sqrt{4\pi\alpha_{b,\tau}(M_Z)}
\frac{1}{\sqrt{1+\tan^2\beta}} v.
\end{array}
\eeqn{Yukawa masses}
Here $v \equiv \sqrt{v_1^2+v_2^2}$. Note that
for fixed mass $m_b$ the bottom coupling $\alpha_b$  increases with
$\tan \beta$, whereas the top coupling increases when $\tan \beta$
diminishes. Evidently, $m_b/m_\tau=\sqrt{\alpha_b/\alpha_\tau}.$

Applying, for
instance, the method of the effective potential of
ref. \cite{Falck}, one finds the RGE for the couplings
$\alpha_x$:
\begin{eqnarray}
\alpha_t' &=&(\sum_i b^i_u\ \alpha_i -6 \alpha_t-
\alpha_b-\alpha_{\nu_\tau}\
\theta_R)\alpha_t \label{top rge}\\
\alpha_b'&=&(\sum_i b^i_d\ \alpha_i -6 \alpha_b-
\alpha_t-\alpha_\tau)\alpha_b
\label{b rge}\\
\ \ \alpha_{\nu_\tau}'&=&(\sum_i b^i_\nu\ \alpha_i
-4 \alpha_{\nu_\tau}\ \theta_R-  \alpha_\tau -3\alpha_t )
\alpha_{\nu_\tau}
\label{nu rge}\\
\alpha_\tau'&=&(\sum_i b^i_e\ \alpha_i -
4 \alpha_\tau-\alpha_{\nu_\tau}\ \theta_R -
3 \alpha_b) \alpha_\tau,\label{tau rge}
\end{eqnarray}
where ${ A}'\equiv {dA/dT}$, $T={1/2 \pi}\ {\rm log}(M_G/Q)$,
and $\theta_R(T)\equiv \theta(T-T_R)$ is the step function
which describes the effect of the $\nu_R$ decoupling at $M_R$
($T_R\equiv 1/2\pi\  \log(M_G/M_R)$).
The gauge coupling constants $\alpha_i$
($i$ = $SU_3, SU_2, U_1$)
enter, in particular, the neutrino RGE with the
coefficients: $b_\nu=(0, 3, {3\over 5}).$
The signs in the eqs. (\ref{top rge}-\ref{tau rge})
reflect the  well known fact that
during the evolution from the high to the low energy scale,
the effect of the gauge coupling
constants is to increase the Yukawa couplings, while the effect
of the Yukawa itself is opposite.

The following preliminary remarks are in order.
Neglecting all the Yukawa coupling but $\alpha_t$ in
(\ref{top rge}) (which is justified for small values of
$\tan\beta$) one gets the equation
\beq
\alpha_t'=(\sum b^i_u\ \alpha_i-6 \alpha_t)\alpha_t.
\eeqn{IR top}
For large values of $\alpha_t$ its solution has  the
well known IR fixed point beha\-viour, which is kept
when the system of equations is taken into account without
simplification. At
large scales $\alpha_t$ diverges; small variations of $\alpha_t(T_Z)$,
and therefore $m_t(M_Z)$
result in strong changes of $\alpha_t$ at the GU scale
(here $T_Z\equiv 1/2\pi\  \log(M_G/M_Z)$).
It is worthwhile to notice that the
running top mass $m_t(M_Z)$ coincides
numerically with
the pole mass $M_t$ up to 2\% if $M_t$ is in the range
$150-200$ GeV.

In fig. 1a-c we show the divergency lines  on the
$m_t(M_Z) - \tan\beta$-
plot which correspond to $\alpha_t(0)=0.5$ at $M_{G}$. They where
found by numerical solution of the system
(\ref{top rge}-\ref{tau rge})
at maximally admitted
value  of $m_b$ which
corresponds to the pole mass $M_b^{pole} = 5.2\ \GeV$.
For values of the parameters $m_t(M_Z) - \tan\beta$
above the divergency  lines the perturbative approach
is invalid, the mass $m_t$ blows up before GU scale.
The increase of gauge coupling $\alpha_s$
results in a decrease of the $\alpha_t$ at large scales,
and consequently, relaxes the divergency bound on
$m_t(M_Z)$ (fig. 1a-c).
The Yukawa couplings have an opposite effect. This
explains the bending of the curves for large values
of $\tan\beta$ ( according to (5)
an increase of $\tan \beta$ corresponds to  increase of the bottom
coupling $\alpha_b$).
Neutrino renormalization also  makes the divergency bound on $m_t$ more
stringent. The corrections due to neutrinos:
$|\Delta m_t|\ltap 5\ \GeV,$
are comparable with the uncertainty related to  $\alpha_s$.
\eject

\noindent{\large 3.
Renormalization of $m_b/m_\tau$ and the $b-\tau$ unification}

We will suggest the $b-\tau$ unification,
so that the $b$ and $\tau$  Yukawa couplings coincide at
$M_G:$
\beq
\alpha_b(0)=\alpha_\tau(0).
\eeqn{b tau unif}
For definiteness we also suggest the equality of the couplings
of the top
quark and the tau neutrino:
\beq
\alpha_t(0)=\alpha_{\nu_\tau}(0).
\eeqn{t nu-tau unif}
In fact, the  equalities (\ref{b tau unif}, \ref{t nu-tau unif})
appear  in GU theories with the  Higgs
multiplets of the lower dimension\footnote{There
are no similar relations  for $1^{st}$ and $2^{nd}$ generations.
One can  suggest that only the particles from the third
generation acquire the masses at the tree level in interactions with
Higgs multiplets. Masses of the lightest generations appear as radiative
corrections or/and from effective nonrenormalizable
terms.}.

Using (\ref{top rge}-\ref{tau rge}) one finds the RGE for the mass ratio:
\beq
\left(\frac{m_b}{m_\tau}\right)'=
\frac{1}{2}\left(\sum_i (b^i_d-b^i_e)\ \alpha_i
 -3 (\alpha_b- \alpha_\tau)-
(\alpha_t-\alpha_{\nu_\tau}\ \theta_R)\right)
\left(\frac{m_b}{m_\tau}\right).
\eeqn{R RGE}
The coefficient $1/2$ reflects that $m\sim \sqrt{\alpha}.$
Note that the effect of down quark and charge lepton is enhanced
by factor 3.
The following conclusions can be drawn from \eq{R RGE} immediately.
$(i)$ The increase of $\alpha_t,$ or of $\alpha_b$
tend to decrease the ratio. This is  a key ingredient to
achieve the $b-\tau$ unification.
$(ii)$ The gauge coupling $\alpha_s$  works in opposite direction:
the predicted value of the $m_b/m_\tau$ ratio
increases with $\alpha_s$.
$(iii)$ The neutrino renormalization  also results in
increase of $m_b/m_{\tau}$. Therefore to achieve the mass
unification one should take larger values of $\alpha_t$ to
compensate the effects of neutrino or/and $\alpha_s$.

These features can be seen in fig. 2 a-c.
The ratio of masses
of $m_b/m_{\tau}$ predicted from the $b-\tau$ unification
is shown  as function of $\alpha_t(0)$
at the GU scale for different values $M_R$ and $\alpha_s$.
The prediction
is compared with the experimental
mass ratio which has been found
by converting the pole masses in running masses, and
running them from low energy scale to $M_Z$.
(We take
into account two-loop SM renormalization effects as  explained in
\cite{florida}).
The two horizontal lines
in fig. 2 correspond to two values of pole mass:
$M_b^{pole}(\max)=4.85\ \GeV$ (being $2\sigma$ above the value
quoted in \cite{mb experimental}) and
$M_b^{pole}(\max)=5.2\ \GeV$ (an extreme upper bound).
As follows from the figure, the
agreement between the predicted and the experimental
(running) values can be achieved only for large values of
$\alpha_t(0)$ and of $\tan\beta$ (\ie large
$\alpha_b$). A decrease of $M_R$ results in increase of the ratio.
The neutrino renormalization effect prevents from
the $b - \tau$ unification at small $\tan\beta$:
$\tan \beta < 10$. If for example $M_R < 10^{-3} M_G,$ no unification
can be achieved for  $\alpha_s > 0.115$ and $\alpha_t$
up to the divergency limit.
The reliable upper bounds on $m_t$ will allow to further
strengthen the bounds on  $\tan\beta$ and $M_R$ with respect to
those from the divergence
of $\alpha_t(0).$
Indeed, taking as an example the case
$\alpha_s(M_Z)=0.12$ and $\tan\beta= 3,$
one gets that the top mass range $m_t=164\div 184\ \GeV$ \cite{CDF}
corresponds to $\alpha_t(0)=0.027\div 0.55$
for $M_R=10^{-6}M_G$ and
$\alpha_t(0)=0.025\div 0.22$ for  $M_R=M_G$.
Consequently, the upper bound on $m_t$ can give an upper
bound on $\alpha_t(0)$ which is appreciably lower than the divergence
limit.

The dependence of the limits for  $\tan \beta$ and $M_R$
on $m_t$ can
be seen in fig. 3. Here the $b-\tau$ unification curves are
shown in the $(m_t(M_Z)-\tan\beta)$--plane. The curves were found
by solving
the system (\ref{top rge}-\ref{tau rge}) and
correspond to the highest possible value  $M_b^{pole}=5.2\ \GeV$.
Evidently a situation for smaller b-masses  is even worser.
As follows from fig. 3 in case of the  decoupled RH neutrinos, $M_R=M_G,$
the unification solution exists in all the range of
the  $\tan \beta$ values. With a decrease of $\tan \beta$
(decrease $\alpha_b$)  the dump effect
of the bottom  Yukawa coupling on $m_b/m_{\tau}$ decreases.
This should be compensated by the increase of the renormalization
effect of the top quark which results
in increase of the top mass. In the case  $M_R \ll M_G$ the
neutrino effects
reduce the influence of the top interaction, and therefore
for a given value of $\tan\beta$ the coupling
$\alpha_t(0)$ has to be larger,
in comparison with the case $M_R=M_G$.
For this reason the unification curves of fig. 3 shift to larger
values of $m_t$. Moreover, for small $M_R$ with decrease of $\tan \beta$
these curves  exit from the perturbative domain. The terminations of
curves at some values of $\tan\beta$
correspond with a good precision to the divergency
limit of fig. 1.

As follows from  fig. 3 for light $M_R$ the
$b - \tau$ unification and the  condition
that $\alpha_t$ is in the perturbative domain
allow one to  get the lower bound on
$\tan\beta,$ and an upper bound on the top mass.
For example, at $M_R = 10^{-6}M_G$ one finds
$\tan \beta \gtap$ 31, 46, 55 and $m_t\ltap$ 188, 186, 184 $\GeV$
for $\alpha_s =$ 0.115, 0.120, and 0.125
correspondingly.

In fig. 4 we show explicitly the lower bounds on $\tan\beta$
and $M_R$
obtained from
the $b-\tau$ unification condition ($M_b^{pole}\leq 5.2\ \GeV$)
and the convergency limit for $\alpha_t$ for different
values of $\alpha_s.$
Let us stress a strong dependence of the bounds on
strong coupling constant.

Evidently, for realistic values of
$M_b$
the conclusions derived from fig. 3 and 4 strengthen. Since
the bottom Yukawa coupling is
lower in this case,
$\alpha_t$ has to be larger, and the unification curves
closer to the divergency curves.
This fact disfavours
strongly the low $\tan\beta$ region also for low value of $\alpha_s.$

Note that
for relatively small values of $\tan\beta$, and consequently, the
small Yukawa coupling of $b$ and $\tau$, it is possible
to write  the  explicit
expression for the mass ratio.
Neglecting the coupling constants $\alpha_b,\ \alpha_\tau$
in the brackets of (\ref{top rge}-\ref{tau rge}), as well as
$\alpha_{\nu_\tau}$
in (\ref{top rge}) one finds the solution of the simplified
system of equations:
\beq
\frac{m_b}{m_\tau}\left(T_Z \right)=
\left( \frac{E_b(T_Z)}{E_{\tau}(T_Z)}\right)^{1/2}
\frac{D_{\nu_\tau}(T_Z)^{1/8}}{D_t(T_Z)^{1/12}}~~,
\eeqn{approx ratio}
where the renormalization effect of the Yukawa couplings
are summarized in the functions $D_t$ and $D_{\nu_\tau},$ while
the four functions $E_x$ describe the  gauge interaction effects:
\beq
\begin{array}{ccl}
E_x(T)&=&\Pi_i \left( \frac{\alpha_i(T)}{\alpha_i(0)}\right)^{-b_x^i/b_i}
\ \ \ \ \ \ \ \ {\mbox{($x=t,b,\tau,\nu$)}}\\
\ E_{\nu_\tau}(T)&=&\frac{E_\nu(T)}{D_t(T)^{1/2}}\\
D_t(T)&=&1+6\ \alpha_t(0)\ \int_0^T E_u(x) dx\\
\ D_{\nu_\tau}(T)&=&1+4\ \alpha_{\nu_\tau}(0)
\ \int_0^T E_{\nu_\tau}(x)\ \theta_R(x)\ dx
\end{array}
\eeqn{auxiliary functions}
The top  and the neutrino  contributions, that is
$D_t(T)$ and $D_{\nu_\tau}(T)$, are related by
eq. (\ref{t nu-tau unif}). The solution
(\ref{approx ratio}-\ref{auxiliary functions})
generalizes the result of ref.
\cite{Ibanez} by including the neutrino effect.

For $\tan \beta < 10 $ the approximate solution \eq{approx ratio}
coincides with
the results of the numerical integration of \eqs{top rge}{tau rge}
within $2\div 3$\%.
Note that according to (15) at small $\tan \beta$  the Yukawa
renormalization
effects do not depend on $m_t$ and on $\tan\beta$ separately:
they enter  the expression only  via  $\alpha_t(0).$
Consequently, for fixed $\alpha_t(0)$ the unification curves do
not depend on $\tan \beta$. At
 $\tan\beta < 10$  the  curves  for different  $\tan\beta$
practically coincide with those
shown in fig. 2 for $\tan\beta = 3$.

{\large 4. Neutrino masses }

Above the $M_R$ scale
eq. (\ref{nu rge}) gives us the RGE of the
see-saw mass of the  tau neutrino, being
$m_{\nu_\tau} = k\ \alpha_{\nu_\tau}$
($k=4\pi v_2^2/ M_R,$
where
$M_R$ the physical mass of the
RH tau neutrino).
Also below this scale
the renormalization of
$k\ \alpha_{\nu_\tau}$ according to \eq{nu rge}
gives us the correct evolution of $m_{\nu_\tau}$
since this evolution coincides with that of the
mass
operator, in the supersymmetric case,
found recently
in \cite{babu chank}.
We will study in a future work the
the relation of this issue
with the non renormalization theorems in supersymmetry.
We include in the
present analysis the
effect of the large neutrino
Yukawa coupling in the
region of momenta above $M_R;$ this results
in large renormalization
effects
for the neutrino masses themselves.

In the context of $b-\tau$ unification the value of $\alpha_t(0),$
and consequently $\alpha_{\nu_{\tau}}(0),$
are calculated once the top mass is
fixed.  Depending on value of
$M_R$ one or two solutions for $\tan\beta$ can
exist (see fig. 3), and consequently there are  one or two solutions
for $\alpha_t(0).$ In our  calculation we  take
the solution with large values of $\tan\beta,$
being interested to have lower values of $M_R.$
For $m_t=174\pm 10 \ \GeV$ and $\alpha_s=0.12$ the
results of numerical solution of system (\ref{top rge}-\ref{tau rge})
can be parameterized by the
following formula
\begin{equation}
m_{\nu_\tau}=\left(10^{+3.5}_{-2.2} \right)\ \eV
   \left(  \frac{10^{12}\ \GeV}{M_R}  \right)
\label{numerical nu mass}
\end{equation}
which holds with a $5\%$ of accuracy
in the range
$M_R=(10^{11}-10^{13})\ \GeV,$ and $15\%$ for $M_R=(10^{10}-10^{14})\ \GeV.$
Previous estimations  of $m_{\nu_\tau}$
in SUSY GUTs  \cite{nu mass} correspond to $M_R$ of the order   $M_G.$
The result (\ref{numerical nu mass}) implies the  $b-\tau$
unification, and the updated  mass range of $m_t$ was used.

The solution for $\alpha_{\nu_\tau}$ can be found explicitly in the
approximation we used to derive the
eq. (\ref{approx ratio}). Moreover, since the eq.
(\ref{nu rge})
does not contain $\alpha_b$ the approximate solution
for $\alpha_{\nu_\tau}$  is reliable even for large values
of $\tan\beta$.
The neutrino mass can be represented
in terms of $m_t(M_Z)$ as:
\begin{equation}
m_{\nu_\tau}=f_G\ f_Y\ \frac{m_t^2}{M_R}
\label{nu mass with corrections}
\end{equation}
where $f_G\equiv E_\nu(T_Z)/E_u(T_Z)$
describes the effect of gauge couplings
renormalization, being $f_G \approx 0.145\mp  0.01$
for $\alpha_s(M_Z)=0.120\pm0.005$. The factor
$f_Y=D_t(T_Z)^{1/2}/D_{\nu_\tau}(T_Z)$ represents
the effect of the top and neutrino Yukawa renormalization.
Confronting  \eq{nu mass with corrections} and (\ref{numerical nu mass})
one finds that the Yukawa renormalization $f_Y$
amounts to a factor larger than
two\footnote{For solutions corresponding to low value of $\tan\beta$
(see fig. 3)
the Yukawa contributions is approximatively doubled.}.
For largest $m_t$ allowed in the model we find
$f_Y\sim 3$ at $M_R=10^{-4} M_G;$
$f_Y$ can be as large as $\sim 8$ for a decoupled neutrino, $M_R=M_G.$

As in the case of the \eq{approx ratio} the Yukawa corrections
depend only on $\alpha_t(0).$

Consider now the effect of flavour mixing. Obviously,
the results are not
changed if the mixing is small both in the Dirac and the Majorana mass
matrices.
In the case of large mixing in the Yukawa matrices,
it is necessary to consider the evolution of the full matrix system.
Defining
the $3\times 3$ hermitian matrices $\alpha_x$:
\beq
\alpha_x={Y_x\cdot Y_x^\dagger \over 4 \pi}\ \ \ \ \ \ \ \
{\mbox{($x=u,d,e,\nu$)}}
\eeqn{matricial alpha}
one can obtain the RGEs by the method of \cite{Falck}.
For example, for the matrix of
the neutrino couplings one has
\beq
 \alpha_\nu'=\sum_i b^i_\nu\ \alpha_i\ \alpha_\nu-\left(3 \alpha_\nu^2+
{1\over 2} \{\alpha_\nu,\alpha_e\}
+{\rm Tr}(3\alpha_u+\alpha_\nu)\alpha_\nu \right)
\eeqn{matrix rge}
Eq. (\ref{matrix rge}) is  valid above the mass scale of the
heavier RH neutrino.
In the case of large mixing the
neutrino renormalization effect
has smaller influence on the third family evolution,
``discharging'' partly on
the light families evolution.
However,  the estimations show that this effect can not ensure
the unification of the lighter down quarks and charged leptons,
being of the order of $15\%$ only.

Another deviation from the one family case can be
related to large mixing in the Majorana sector.
Consider  two
heaviest generations with Dirac mass matrix of the Fritzsch type,
and the  off-diagonal Majorana mass matrix $\hat M_R.$
In this case one gets in lowest order:
\begin{equation}
m_{\nu_\tau}=2 \frac{m_{33} m_{23}}{M_R}
\label{fritzsch}
\end{equation}
(and $\tan\theta\sim m_{23}/m_{33}$), where $m_{ij}$ are
the elements of the Dirac mass matrix. According to (\ref{fritzsch})
the light mass is suppressed by a factor
$m_{23}/m_{33}$ in comparison with the case of diagonal
matrix  $\hat M_R$.
The scale  $M_R$ should be $m_{33}/m_{23}$ times smaller
to get the same value of the light
neutrino mass. Evidently, with diminishing $M_R$ the
renormalization effects
due to neutrino increase.

{\large 5. Discussion and conclusions}

\noindent 1. In
the minimally extended supersymmetric standard model we have found
the relation between $m_b/m_\tau,$ $\alpha_s,$ $\tan\beta,$
$m_t$ and $M_R.$ For masses
$M_R$ being $5-6$ orders of magnitude smaller
than $M_G$ the renormalization effect due
to the neutrino Yukawa couplings
results in an increase of the predicted value
of $m_b/m_\tau$ up to  $10-15 \%.$
In a sense the neutrino renormalization
compensates the effect of top quark
interactions, which make it possible the $b-\tau$ unification
at small $\tan\beta.$ Neutrino renormalization effects
disfavour the $m_b-m_\tau$
unification at small $\tan\beta.$

For fixed values of $\alpha_s$ and  $M_b,$ the $b-\tau$ unification
gives the lower bounds on $\tan\beta$ and $M_R$ (fig. 4).
The smaller $M_R,$ the larger $\tan\beta.$ At
$M_R$ in the intermediate
scale one gets $\tan\beta\gtap 43$ for $\alpha_s>0.115$ and
$M_b\leq 5.2\ \GeV.$
Note that large values of $\tan\beta$ are especially interesting from the
point of view of the unification of all Yukawa couplings as well as
possible relations of the Yukawa and the
gauge coupling.

\noindent 2. The
bounds on $\tan\beta$ and $M_R$ are very sensitive
to values of $\alpha_s$ and $M_b$ used as an input. The increase
of $\alpha_s$ up to the value, \eg $0.125$ (which is suggested
by LEP data \cite{LEP}
and the unification of gauge couplings)
increase  the lower bound on $\tan\beta$ up to $\tan\beta\gtap 56$
for $M_b^{pole}\leq 5.2 \ \GeV.$ We used this conservative bound on
$M_b^{pole}$
which, in a sense, includes possible threshold effects (evaluated at
$\sim 10\%$). The value $M_b^{pole}=4.7\pm 0.05\ \GeV$ implied by
\cite{mb experimental} gives very strong bounds on $\tan\beta$ and
$M_R$ even for small values of $\alpha_s.$ Evidently, future
improvements of the determination of $\alpha_s$ and of $M_b$
will allow to strengthen the conclusion.

\noindent 3. According to the see-saw mechanism $M_R$ determines  masses
of the light neutrinos (or give an upper bound on these masses
if the matrix $\hat M_R$ is appreciably non diagonal).
Future measurements of the tau neutrino mass
will allow to confirm the possibility
of $M_R$ being at the intermediate scale.
In particular, an  observation of the oscillation effects
in the experiments CHORUS and NOMAD \cite{osc}
will strongly suggest the $\nu_\tau$-mass in the
cosmologically interesting domain.
Also the confirmation of the MSW solution of the solar neutrino problem
will testify for $M_R$ in the $10^{10}-10^{12}$ GeV range.

\noindent 4. The violation of the bounds on $\tan\beta$ and  $M_R$ may
testify for the existence of new particles below the GU scale,
some other mechanism of neutrino mass generation, or the absence of the
$b-\tau,$ or $t-\nu_\tau$ unification.

In conclusion, the existence of the RH neutrinos at the intermediate
scale (which is implied in particular by the see-saw mechanism
and the existing hints from studies of solar, atmospheric and
relic neutrinos)
disfavours the $b-\tau$ unification, especially at low $\tan\beta.$
The $b-\tau$ unification gives the lower bounds on $M_R$ and $\tan\beta.$
The bound can be strengthened by further
refining the determination of $\alpha_s$
and of $M_b.$

{\large Acknowledgments}

F.V. would like to thank S. Bertolini for computational help,
Y. Rizos,
M. Cobal and C. Pagliarone for useful discussions.

\end{document}